\documentclass[a4paper]{article}

\usepackage{INTERSPEECH2022}
\usepackage[ruled,vlined]{algorithm2e}
\usepackage{multirow}
\usepackage{xcolor,soul}
\usepackage{adjustbox}
\usepackage{url}
\usepackage{kotex}
\usepackage{subcaption}
\usepackage{amsmath}

\title{Bunched LPCNet2: Efficient Neural Vocoders Covering Devices\\from Cloud to Edge}
\name{Sangjun Park$^1$, Kihyun Choo$^1$, Joohyung Lee$^1$, Anton V. Porov$^2$, Konstantin Osipov$^2$, June Sig Sung$^3$}
\address{
  $^1$Samsung Research, Samsung Electronics, Republic of Korea\\
  $^2$PDMI RAS, Russia\\
  $^3$Mobile eXperience Business, Samsung Electronics, Republic of Korea}
\email{\{sj0.park, khchoo, jooooh.lee\} @samsung.com}
\begin{document}

\maketitle
\begin{abstract}
Text-to-Speech (TTS) services that run on edge devices have many advantages compared to cloud TTS, e.g., latency and privacy issues. However, neural vocoders with a low complexity and small model footprint inevitably generate annoying sounds. This study proposes a Bunched LPCNet2, an improved LPCNet architecture that provides highly efficient performance in high-quality for cloud servers and in a low-complexity for low-resource edge devices. Single logistic distribution achieves computational efficiency, and insightful tricks reduce the model footprint while maintaining speech quality. A DualRate architecture, which generates a lower sampling rate from a prosody model, is also proposed to reduce maintenance costs. The experiments demonstrate that Bunched LPCNet2 generates satisfactory speech quality with a model footprint of 1.1MB while operating faster than real-time on a RPi 3B. Our audio samples are available at \url{https://srtts.github.io/bunchedLPCNet2}.

\end{abstract}

\section{Introduction}


Nowadays, as many types of edge devices emerge, people utilize text-to-speech (TTS) services in their daily life without device-constraints. Although most TTS systems now have been launched on cloud servers, running on edge devices resolves significant concerns, such as latency, privacy, and internet connectivity issues. Note that edge devices include high-resource devices, such as smartphones, and low-resource devices, such as smart watches, wireless earphones, and home assistant devices. To deploy TTS services on these devices optimizing the performance within a given resource is crucial. Additionally, expanding the service capability while maintaining the architecture is worthwhile instead of developing new architectures for specific devices. For this purpose, FBWave~\cite{wu2020fbwave} proposed a flow-based scalable vocoder architecture to easily control the computational cost. However, it demonstrated an insufficient speech quality in cloud servers, even in high-quality mode. This study introduces a vocoder family that can provide a wide coverage on computational cost constraints from cloud to low-resource edge devices.


Many neural vocoders have been proposed since WaveNet, which predicts a waveform directly, was presented~\cite{oord2016wavenet}. Early neural vocoders are based on a computationally expensive auto-regressive (AR) manner~\cite{oord2016wavenet,kalchbrenner2018efficient}, impractical for real-time services. Inferring in parallel on GPUs have also been attempted~\cite{prenger2019waveglow,kumar2019melgan,kong2020hifi}, but few edge devices have a GPU.
On the other hand, LPCNet~\cite{valin-lpcnet} is a light-weight vocoder based on AR architecture that is suitable for on-device inference. To the best of our knowledge, LPCNet is still one of the best candidates that can efficiently generate speech on a CPU, even with a high quality. Many LPCNet variants have been introduced: iLPCNet~\cite{hwang2020improving} and Full-Band LPCNet~\cite{matsubara2021full} were studied to improve speech quality; and Bunched LPCNet~\cite{bunched}, Gaussian LPCNet~\cite{popov2020gaussian}, and Sub-band LPCNet~\cite{cui2020efficient} were studied to reduce complexity without sacrificing speech quality. However, the variants are unsatisfactory for low-resource edge devices in that they target high-end devices, such as smart phones. A quality drop is also observed by shrinking the network capacity for low-resource devices.

In this paper, we introduces Bunched LPCNet2, which is an improved Bunched LPCNet that addresses the problems arising when covering a wide range of devices, from cloud servers to wearable devices such as smart watches and AR/VR glasses. Our contributions include three methods:

\begin{enumerate}
\item Single logistic output layer that extends the coverage of Bunched LPCNet to low-resource constraints.
\item Dual-rate LPCNet for computational and maintenance cost reduction.
\item Insightful tricks that significantly reduce the model footprint.
\end{enumerate}

We present a brief overview of Bunched LPCNet in Section~\ref{section:bunched}, followed by an in-depth description of the proposed methods in Section~\ref{section:proposed}. Section~\ref{section:evaluation} summarizes the evaluation results in terms of computational cost and speech quality. Finally, Section~\ref{section:conclusion} concludes the paper.





\section{Bunched LPCNet}
\label{section:bunched}

\begin{figure}[t]
\centering
\includegraphics[width=0.40\textwidth]{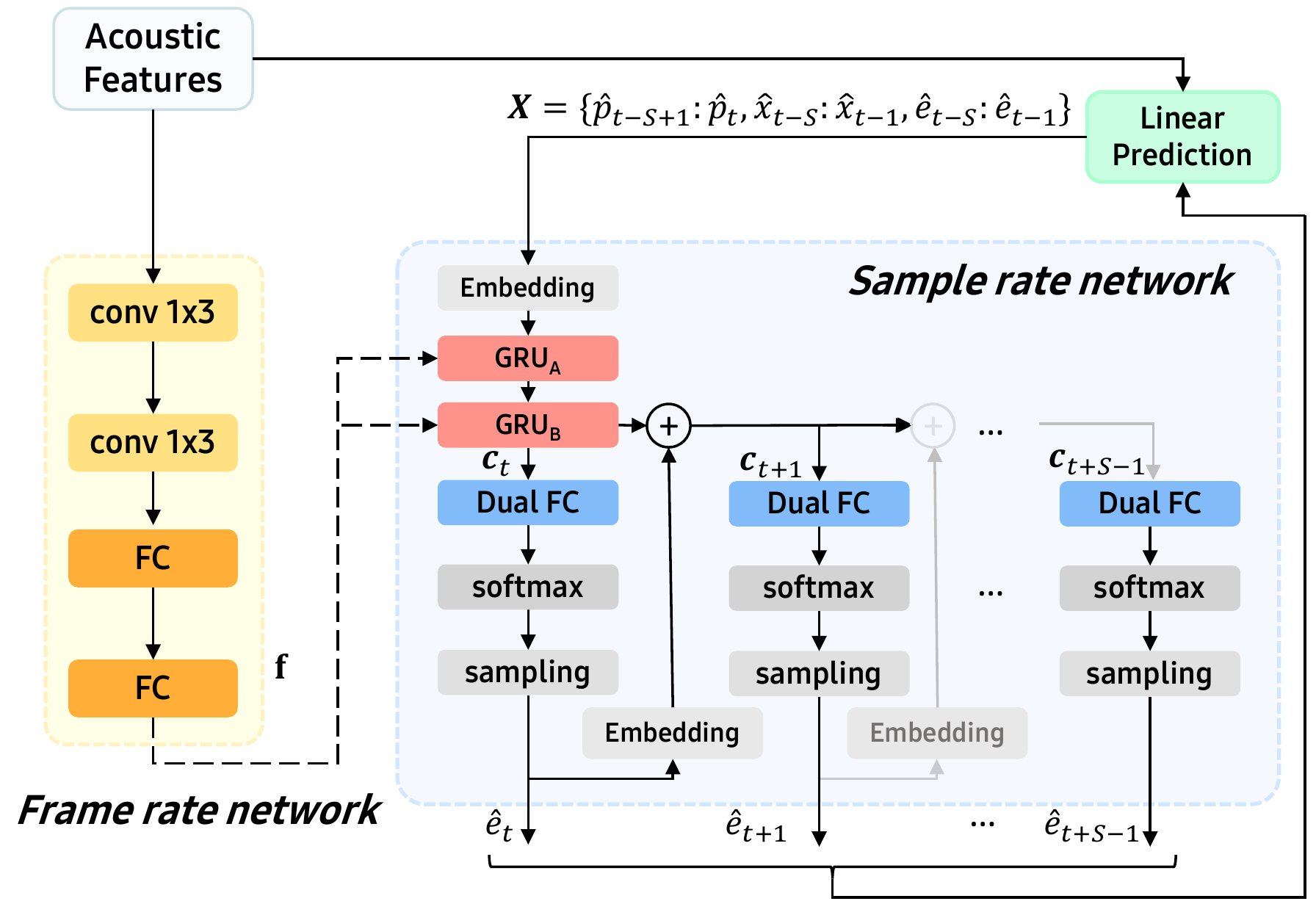}
\caption{Bunched LPCNet overview}
\label{fig:bunched}
\end{figure}

In our previous study~\cite{bunched}, Sample Bunching was proposed as depicted in Figure~\ref{fig:bunched}. It allows LPCNet to generate multiple samples per GRU inference. Bit Bunching was also proposed to reduce the computations in the DualFC and softmax layers by splitting the output bits in two: the higher 7 bits for coarse prediction and lower 4 bits for fine correction.

Bunching techniques are sufficiently efficient for application in mobile devices, e.g., a smartphone. However, several limitations exist for further reducing computations for low-resource devices. (1) Sample Bunching reduces the computations of GRUs by $1/S$ times. This indicates that as the bunch size $S$ increases, the amount of computational reduction is saturated, especially where $S>4$. (2) Reducing the number of $\text{GRU}_{\text{A}}$ units with Bunching techniques significantly degrades the speech quality. Note that $\text{GRU}_{\text{A}}$ is the most computationally expensive layer in LPCNet. (3) The model footprint becomes large because of the embedding tables corresponding to the feedback samples.

To overcome these problems, we improved the Bunched LPCNet architecture with the three methods described in Section~\ref{section:proposed}.


\section{Proposed Methods}
\label{section:proposed}
\subsection{Single logistic output layer}
\label{subsec:sl}

\begin{figure}[b]
    \centering
    \begin{subfigure}[b]{0.23\textwidth}
        \centering
        \includegraphics[width=\textwidth]{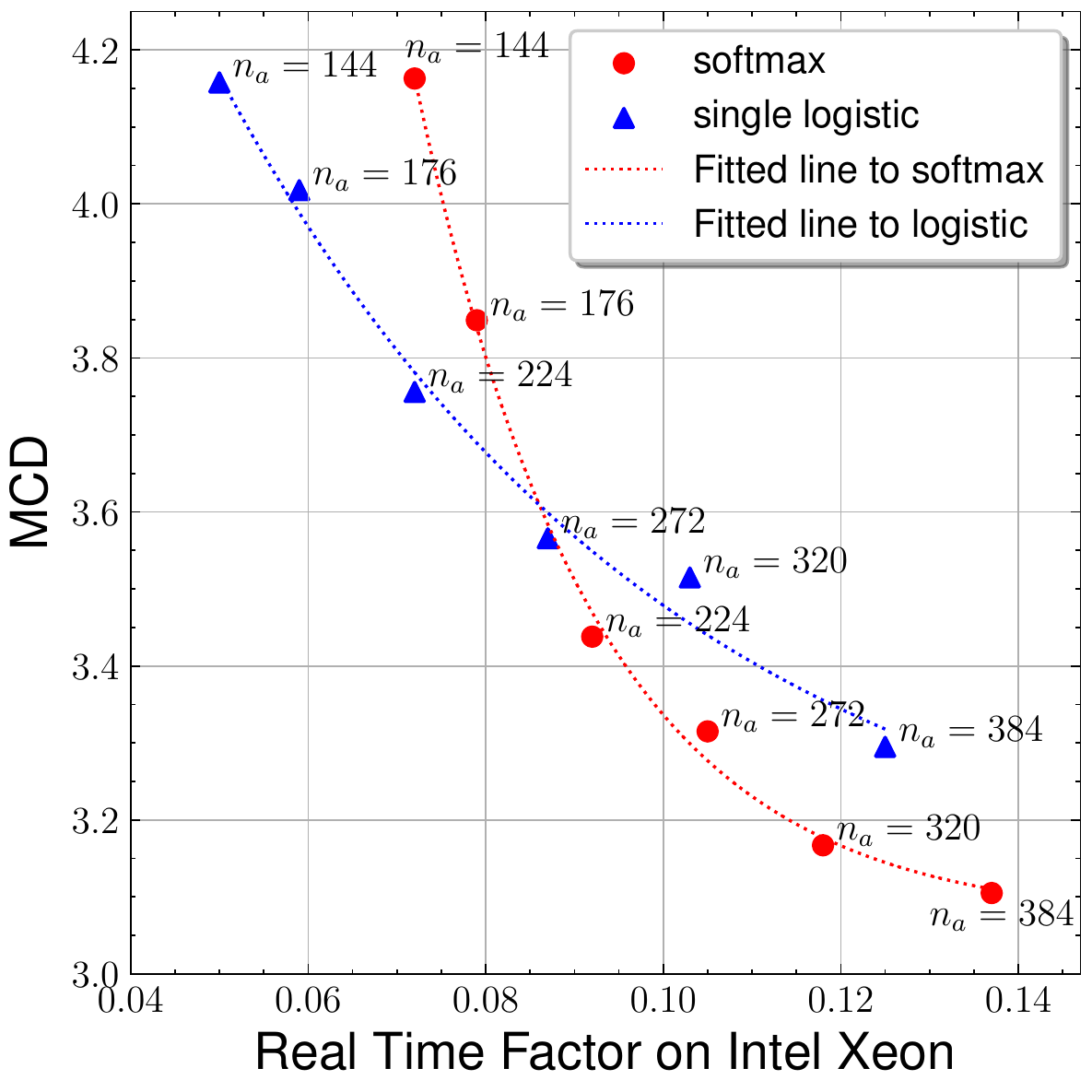}
        \caption{on $\text{GRU}_{\text{A}}$ units ($S$=1) }
    \end{subfigure}
    \hfill
    \begin{subfigure}[b]{0.23\textwidth}
        \centering
        \includegraphics[width=\textwidth]{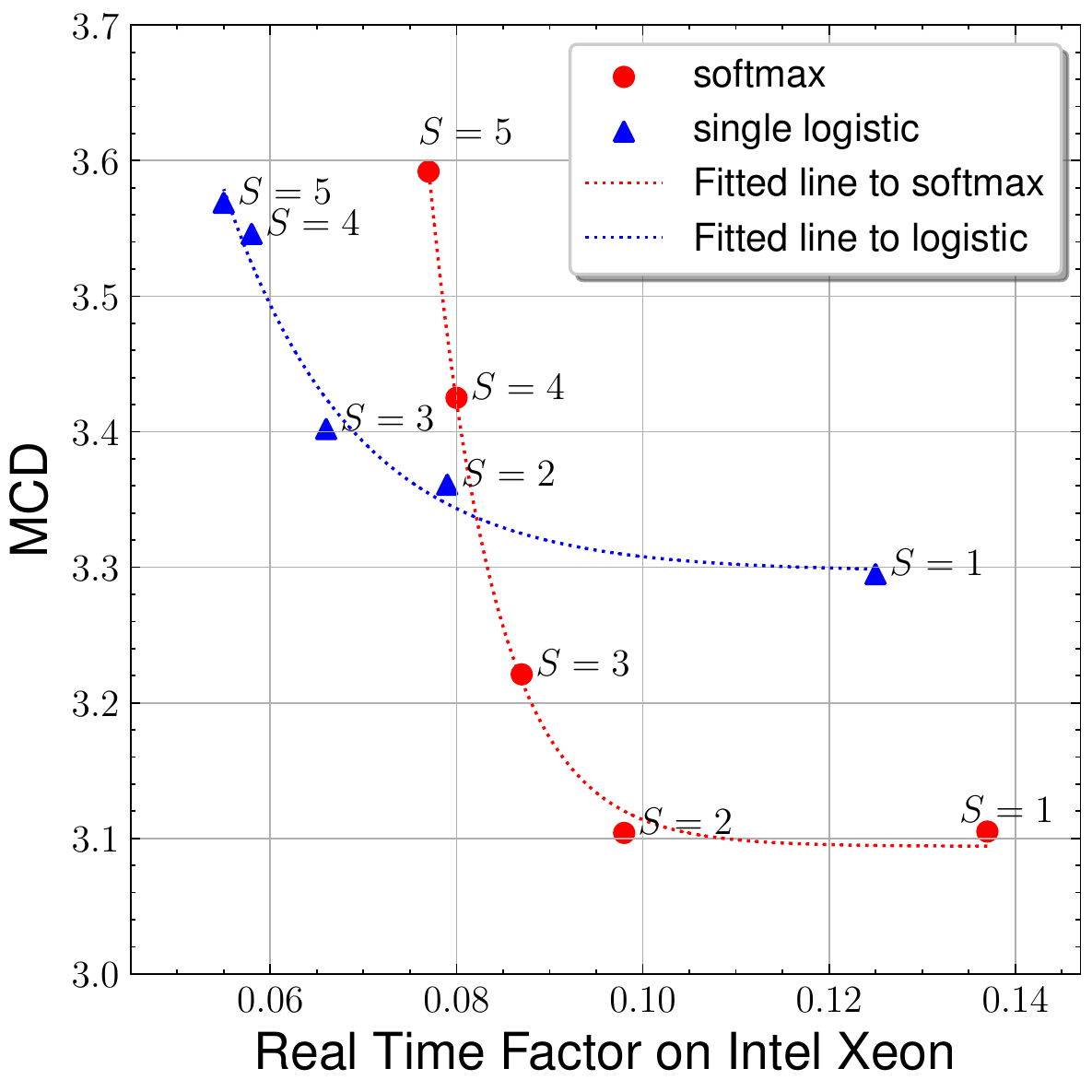}
        \caption{on bunch size ($n_{a}$=384)}
    \end{subfigure}
    \hfill
    \caption{MCD plots of softmax and single logistic output layer}
    \label{fig:mcd}
\end{figure}

Auto-regressive vocoders generate waveforms by sampling from the probability density function with non-parametric or parametric distributions. Non-parametric methods, e.g., a softmax output layer, can model complex distributions and show a good performance in modeling waveforms~\cite{oord2016wavenet, valin-lpcnet}, but the computations in the output layer and sampling process from a categorical distribution are expensive by a large output dimension; for example, 256 for an 8-bits $\mu$-law quantization~\cite{mulaw-quantization}. Additionally, its speech quality is significantly degraded when the model has insufficient capacity.

Accordingly, we attempted to employ parametric distributions based on previous studies~\cite{tacotron2,pmlr-v80-oord18a}, which used a mixture of logistics (MoL) as an output distribution, and determined that a logistic output layer, particularly with single mixture, is a good alternative in terms of efficiency. To demonstrate the efficiency of the logistic layer in complexity reduction, we compare the mel-cepstral distances (MCD)~\cite{mcd} of single logistic (SL) to the softmax output layer with the smaller bunch size $S$ and $\text{GRU}_{\text{A}}$ units $n_{a}$ in Figure~\ref{fig:mcd}. Note that $S=1$ and $n_{a}=384$ correspond to the original LPCNet. As the model capacity decreases, the SL layers work more efficiently than the softmax layers. We believe that a non-parametric loss could lead the network to predict complex distributions beyond the model capacity. That is, it could cause an under-fitting problem when the model capacity is too small. In contrast, a unimodal parametric distribution simplifies an objective and lowers the training difficulty by reducing the burden on the network. It can mitigate the quality reduction despite an insufficient capacity. 

\begin{figure}[t]
\centering
\includegraphics[width=0.25\textwidth]{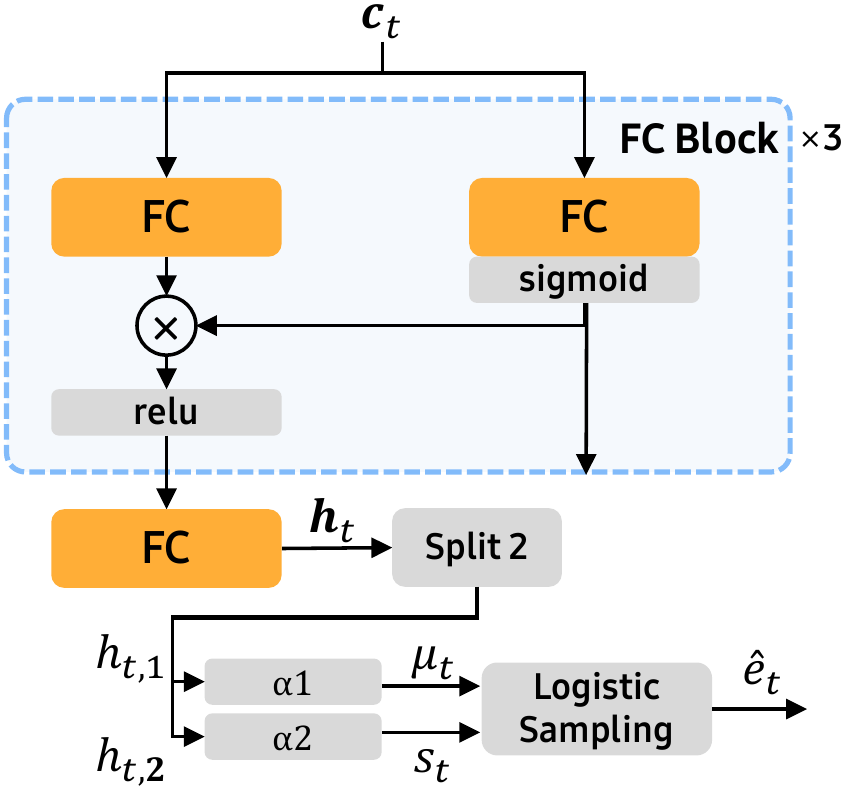}
\caption{Network architecture for single logistic output layer}
\label{fig:sl}
\end{figure}

Figure~\ref{fig:sl} presents our network architecture for the SL layer. The FC block comprises fully-connected (FC) layers with 16 units, and the output dimension of a final FC layer is 2. A location parameter $\mu_{t}$ and scale parameter $s_{t}$ are obtained from the custom activation functions $\alpha 1$ and $\alpha 2$ as follows.
\begin{align}
    \label{eq1}
    \mu_{t} &= \alpha 1(h_{t,1}) = \tanh{(h_{t,1}/u)} \\
    \label{eq2}
    s_{t} &= \alpha 2(h_{t,2}) = \exp{(\tanh{(h_{t,2})} \cdot v - w)}
\end{align}
$u$, $v$, and $w$ help maintain the training stable, and are set empirically, to 64, 16 and 6, respectively. The model is trained to maximize the discretized logistic likelihood~\cite{pixelcnn} on normalized 16-bit samples from -1 to +1. Finally, $\hat{e}_{t}$ is sampled from the logistic distribution, as defined in (\ref{eq3}).
\begin{equation}
    \hat{e}_{t} = \mu_{t} + T \cdot s_{t} \cdot \ln(\epsilon/(1-\epsilon)), \quad \epsilon \sim \text{Uniform}(0,1)
    \label{eq3}
\end{equation}
where $T$ denotes a temperature parameter. When $\hat{e}_{t}$ is fed to the embedding tables, the 8-bit $\mu$-law quantization is applied.

Employing the deep and narrow layers, the proposed architecture demonstrates a better performance while satisfying a lower complexity than the softmax layer. Moreover, the sampling process of the SL layer is undoubtedly simple and computationally cheap using only scalar operations.


\begin{table*}[t]
	\caption{System Configurations}
	\label{tab:systems}
	\centering
	\begin{tabular}{ c|ccccccc }
		\toprule
		\textbf{Systems} & \textbf{Output Layer} & $\bm{S}$ & $\bm{n_{a}}$ & $\bm{n_{e}}$ & $T$ & \textbf{Sampling rate} & \textbf{Target device}\\
		\midrule
		\textit{B-LPCNet}\cite{bunched} & Softmax with Bit Bunching & 4 & 384 & 128 & 0.75 & 24 kHz & High-end \\
		\midrule
		\textit{B-LPCNet2-L} & Softmax & 1 & 384 & 1 & 0.75 & 24 kHz & Cloud \\
		\textit{B-LPCNet2-R} & Single logistic & 2 & 224 & 1 & 0.75 & 24 kHz & High-end \\
		\textit{B-LPCNet2-S} & Single logistic & 5 & 176 & 1 & 0.65 & 24 kHz & Low-end \\
		\textit{B-LPCNet2-S16} & Single logistic & 5 & 176 & 1 & 0.65 & 16 kHz & Low-end \\
		\bottomrule
	\end{tabular}
\end{table*}

\subsection{DualRate LPCNet}
\label{subsec:dualrate}

For low-resource devices, a more compact model would be required. However, a model with low capacity inevitably generates annoying sounds, e.g., noisy or muffled sounds. Reducing the sampling rate could be a solution of this problem.

\begin{figure}[b!]
\centering
\includegraphics[width=0.48\textwidth]{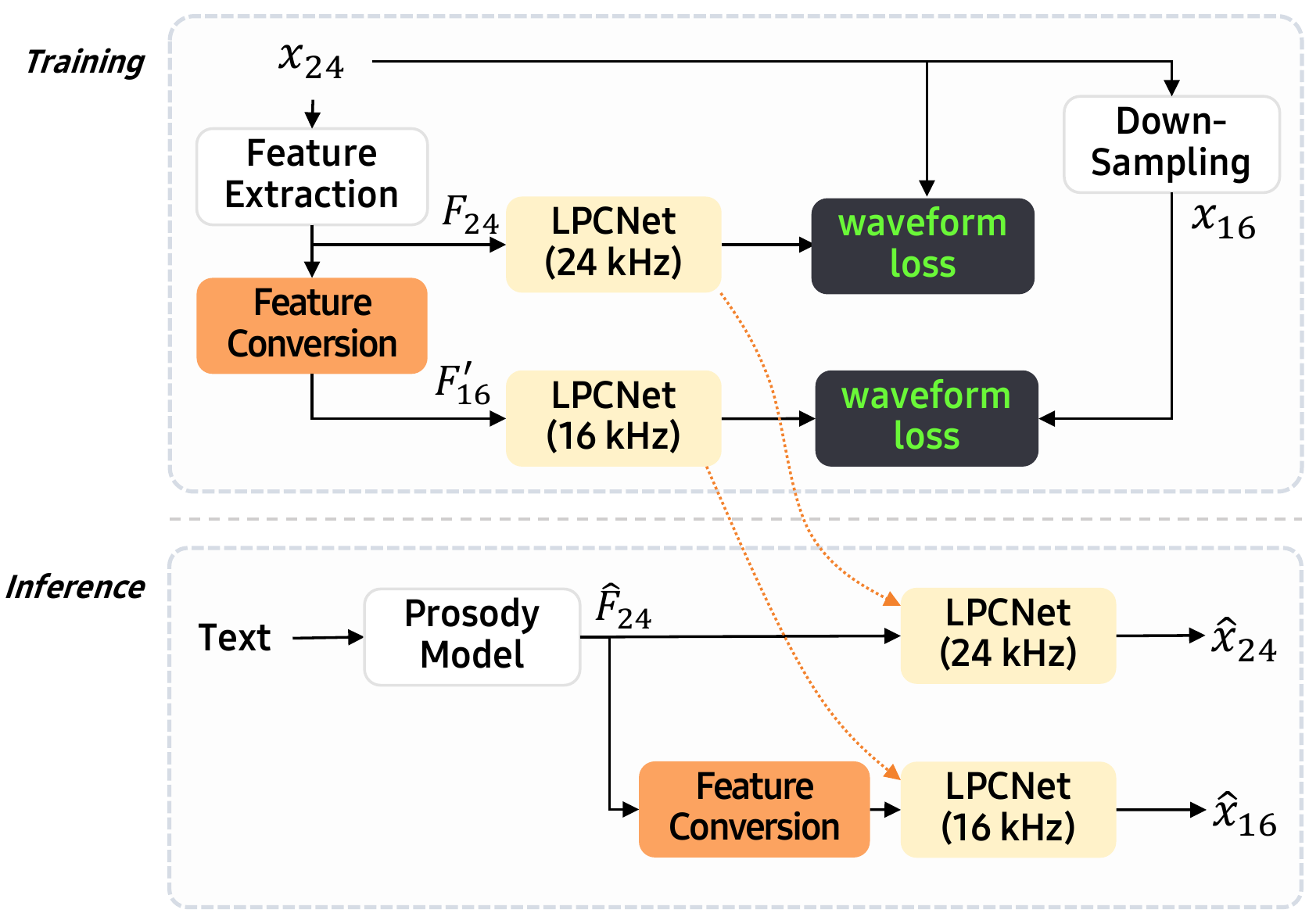}
\caption{Training / inference procedure of DualRate LPCNet.}
\label{fig:dualrate}
\end{figure}

By the way, a prosody model of a TTS system requires continuous maintenance, even after deployed, so that the pronunciation errors or unnatural prosody for given texts can be fixed. In other words, employing an additional prosody model for a low sampling rate vocoder leads to a maintenance cost increase. For this reason, we propose a DualRate LPCNet, which is an architecture that provides two sampling rates (24 kHz and 16 kHz) from one prosody model, as depicted in Figure~\ref{fig:dualrate}. Note that $x_{i}$ and $F_{i}$ denote the $i$ kHz waveform and acoustic features extracted from $x_{i}$, respectively.

The feature conversion block computes $F'_{16}$, which is the input of 16 kHz LPCNet, to remove unnecessary information from $F_{24}$ and obtain higher mutual information with a target $x_{16}$.
The acoustic features $F_{24}$, which comprise 20 cepstral coefficients, a pitch period, and a pitch correlation, are converted as follows.
\begin{itemize}
    \item The 20 cepstral coefficients are extracted with 20 CELT bands from the Opus codec~\cite{opus}. The 20 CELT bands cover up to a 12 kHz bandwidth, and the last two bands cover from 8 kHz to 12 kHz. The two unnecessary bands are removed by sequentially computing the IDCT, low-pass filter, and DCT. Finally, the 18 cepstral coefficients that cover up to an 8 kHz bandwidth are obtainable.
    \item The pitch period is multiplied by the ratio of sampling rates $(\frac{16k}{24k}=\frac{2}{3})$.
    \item The pitch correlation is used as is.
\end{itemize}
Using the 16 kHz LPCNet trained with the converted $F'_{16}$, we can save on both computational and maintenance costs. Similarly, the TTS systems for given devices can be easily deployed by choosing suitable LPCNet models with various complexities.


\subsection{Optimizing model footprint}
\label{subsec:embed}

Although Sample Bunching has an efficient architecture for reducing the computational complexity, it increases the number of model parameters by the embedding tables corresponding to the feedback samples $X$, where $\bm{X}=\{\hat{p}_{t-S+1}:\hat{p}_{t}, \hat{x}_{t-S}:\hat{x}_{t-1},\hat{e}_{t-S}:\hat{e}_{t-1} \}$, as depicted in Figure~\ref{fig:bunched}. As mentioned in~\cite{bunched}, $E_{x \in \bm{X}}$ is converted into a pre-computed lookup table $E'_{x}$ obtained from the multiplication of $E_{x}$ and $U_{x}$, the input weight matrix in $\text{GRU}_{\text{A}}$, to replace matrix-vector multiplication operations with vector-vector addition operations. Specifically, $E_{x}$ with a $[2^{8}, n_{e}]$ shape is converted to $E'_{x}$ with a $[2^{8}, 3 \cdot n_{a}]$ shape, where $n_{e}$ and $n_{a}$ denote the embedding dimension of $E_{x}$ and number of $\text{GRU}_{\text{A}}$ units, respectively. The total number of parameters in $E'_{x}$ is $2^{8} \cdot 3 \cdot n_{a} \cdot (3 \cdot S)$. With the default hyperparameters of Bunched LPCNet, $n_{e}=128$ and $n_{a}=384$, their parameters account for 70\% of the model, even when $S=1$. Therefore, we focus on reducing the embedding table size.

To optimize the model footprint, we attempted to feedback the scalar values into $\text{GRU}_{\text{A}}$. The first method worth considering is to feedback the scalar values directly without the embedding table. However, this wastes the model capacity to transform the scalar values into the desired representation from the network. To feedback desired values into the network without wasting model capacity, a nonlinear mapping function from a quantized value to a continuous value would be useful, which is the embedding table with $n_{e}=1$. This could degrade the speech quality compared to when $n_{e}=128$. However, we determined that it has a small effect on the performance, whereas the size of $E_{x}$ and $U_{x}$ can be significantly reduced.

Note that $n_{e}$ does not affect the size of the dumped model owing to $E'_{x}$ with the fixed shape $[2^{8}, 3 \cdot n_{a}]$ being independent from $n_{e}$. Thus, we dumped $E_{x}$ with the $[2^{8}, n_{e}]$ shape and $U_{x}$ with the $[n_{e}, 3 \cdot n_{a}]$ shape separately into a model instead of $E'_{x}$ (hereafter referred to as separated format). Note that it works only when $2^{8} \cdot n_{e} + n_{e} \cdot 3 \cdot n_{a} < 2^{8} \cdot 3 \cdot n_{a}$. For example, when $n_{e}=128$, $n_{a}$ must be greater than 85. The greater $n_{a}$ and the less $n_{e}$ achieve the higher reduction ratio. The total number of parameters is reduced to $(2^{8} \cdot n_{e} + n_{e} \cdot 3 \cdot n_{a} ) \cdot (3 \cdot S)$ without performance degradation. $E'_{x}$ can be restored during inference engine initialization. The separated format facilitates an extremely small model footprint: 70\% reduction for $S=1$ and 82\% reduction for $S=2$ when $n_{e}=1$ and $n_{a}=384$.

\section{Experimental Results}
\label{section:evaluation}
\subsection{Experimental environment}
The several combinations of hyper parameters were chosen empirically to satisfy various device specifications as summarized in Table~\ref{tab:systems}, compared with Bunched LPCNet. We attempted to control the complexity of the models using $S$ and $n_{a}$ and determined an output layer to efficiently work for a given model capacity. For example, \textit{B-LPCNet2-R} was chosen because the SL layer is more efficient than the softmax layer where $S=2$ and $n_{a}=224$ in Figure~\ref{fig:mcd}. In addition, the temperature parameter $T$ was adjusted to suppress the noise generated by the low-complexity models. Finally, the \textit{B-LPCNet2-L}, \textit{B-LPCNet2-R}, and \textit{B-LPCNet2-S} target cloud, high-end, and low-end devices, respectively, and \textit{B-LPCNet2-S16} was prepared for low-resource devices that need to lower the  sampling rate. The other hyperparameters were set identically to~\cite{bunched}.

A Tacotron variant architecture~\cite{lpctron} was employed as a prosody model to evaluate the speech quality in the TTS pipeline. The systems were trained using two datasets: one of a professional English male speaker (17-hours with 10,000 utterances) and another of a professional English female speaker (15-hours with 7,612 utterances). Each vocoder model and prosody model were trained with a single speaker. Sixty utterances were used as a test set, and one percent of the remaining utterances were used as the validation set for training.

\subsection{Performance}
For the quality evaluation, we conducted mean opinion score (MOS) tests~\cite{itu-t} on the Amazon Mechanical Turk platform with 300 people and 120 unseen test utterances. For the complexity evaluation, we measured the RTF (Real Time Factor) on two devices: 1) an AWS c5.4xlarge instance (Intel Xeon Platinum 8124M CPU @ 3.00GHz) with Ubuntu 18.04 - representing a cloud device and 2) a Raspberry Pi (RPi) 3B v1.2 (BCM2837 @ 1.20 GHz) with Tizen 6.5 - representing an edge device. Our implementation was optimized using SIMD (Single Instruction Multiple Data) with single thread for each architecture. To demonstrate that our systems are as efficient as conventional vocoders, we also measured the complexity of the WORLD vocoder~\cite{world}\footnote{An open source code at \url{https://github.com/mmorise/World} was used }. Table~\ref{tab:mos} presents the evaluation results.

\begin{table}[t]
	\caption{MOS with 95\% confidence intervals and real time factor on each CPU architecture}
	\label{tab:mos}
	\centering
	\begin{tabular}{ c|ccc }
		\toprule
     	\multirow{2}{*}{\textbf{Systems}} &
		\multirow{2}{*}{\textbf{MOS}} &
		\multicolumn{2}{c}{\textbf{RTF}} \\
		& & \textbf{Intel Xeon} & \textbf{RPi} \\
     	\midrule
		\textit{Original(24kHz)} & $4.51 \pm0.04$ & - & - \\
		\textit{Original(16kHz)} & $4.12 \pm0.05$ & - & -\\
		\midrule
		\textit{B-LPCNet} & $3.95 \pm0.06$ & 0.072 & 2.394 \\
		\midrule
		\textit{B-LPCNet2-L} & $4.08 \pm0.06$ & 0.137 & 5.538 \\
		\textit{B-LPCNet2-R} & $3.99 \pm0.06$ & 0.051 & 1.657 \\
		\textit{B-LPCNet2-S} & $3.90 \pm0.06$ & 0.030 & 0.720 \\
		\textit{B-LPCNet2-S16} & $3.81 \pm0.06$ & 0.021 & 0.507 \\
		\midrule
		\textit{WORLD} & - & 0.075 & 1.808 \\
		\bottomrule
	\end{tabular}
\end{table}

The \textit{B-LPCNet2} systems with an SL layer demonstrated a highly efficient performance. The \textit{B-LPCNet2-L} generated sufficiently high fidelity speech for cloud TTS. The \textit{B-LPCNet2-R} achieved a lower complexity than the \textit{B-LPCNet} and even the WORLD vocoder. The \textit{B-LPCNet2-S} and \textit{B-LPCNet2-S16} works 7.7x and 10.9x faster than the \textit{B-LPCNet2-L}, respectively, on the RPi while maintaining a satisfactory speech quality. Employing them with a lightweight prosody model, a real-time TTS system is deployable on low-resource edge devices.

\subsection{Model footprint}

To investigate the efficiency in terms of the model footprint, we compared the MOS results to the model size. Notably, the separated format in Section~\ref{subsec:embed} is applied to all evaluated systems, and the model size included file headers less than 1 KB. Table~\ref{tab:footprint} summarize the results.

\begin{table}[h]
	\caption{MOS with 95\% confidence intervals and the model footprint for different embedding dimensions}
	\label{tab:footprint}
	\centering
	\begin{tabular}{ l|ccc }
		\toprule
		\multicolumn{1}{c|}{\textbf{Systems}} & \textbf{MOS} & \textbf{Model Size} \\
     	\midrule
     	\textit{B-LPCNet} & $3.95 \pm0.06$ & 10.141 MB \\
     	\midrule
     	\textit{B-LPCNet2-L} & $4.08 \pm0.06$ & 1.136 MB \\
     	~└ $n_{e}=128$ & $4.07 \pm0.06$ & 3.496 MB \\
     	\midrule
     	\textit{B-LPCNet2-R} & $3.99 \pm0.06$ & 1.135 MB \\
     	~└ $n_{e}=128$ & $3.97 \pm0.06$ & 3.834 MB \\
     	\midrule
     	\textit{B-LPCNet2-S} & $3.90 \pm0.06$ & 1.099 MB \\
     	~└ $n_{e}=128$ & $3.87 \pm0.06$ & 4.898 MB \\
     	\midrule
     	\textit{B-LPCNet2-S16} & $3.81 \pm0.06$ & 1.071 MB \\
		\bottomrule
	\end{tabular}
\end{table}

The systems with $n_{e}=128$, including \textit{B-LPCNet}, has a relatively large number of parameters. This results from the model size of Bunched LPCNet highly depending on the bunch size, embedding dimension, and $\text{GRU}_{\text{A}}$ units, as mentioned in Section ~\ref{subsec:embed}. In contrast, the \textit{B-LPCNet2} architectures show significantly small footprints of approximately 1.1 MB. Indeed, without the separated format, the size of the \textit{B-LPCNet} is 15.4 MB. Compared to the original Bunched LPCNet, this means that the \textit{B-LPCNet2-R} generates higher speech quality with only 7\% of the parameters.

To confirm the effect of the embedding dimension $n_{e}$ on speech quality, we also compared the MOS results for $n_{e}=1$ and $n_{e}=128$. We expected that the speech quality would degrade considerably, but the quality difference is negligible.

\begin{figure}[h]
\centering
\includegraphics[width=0.4\textwidth]{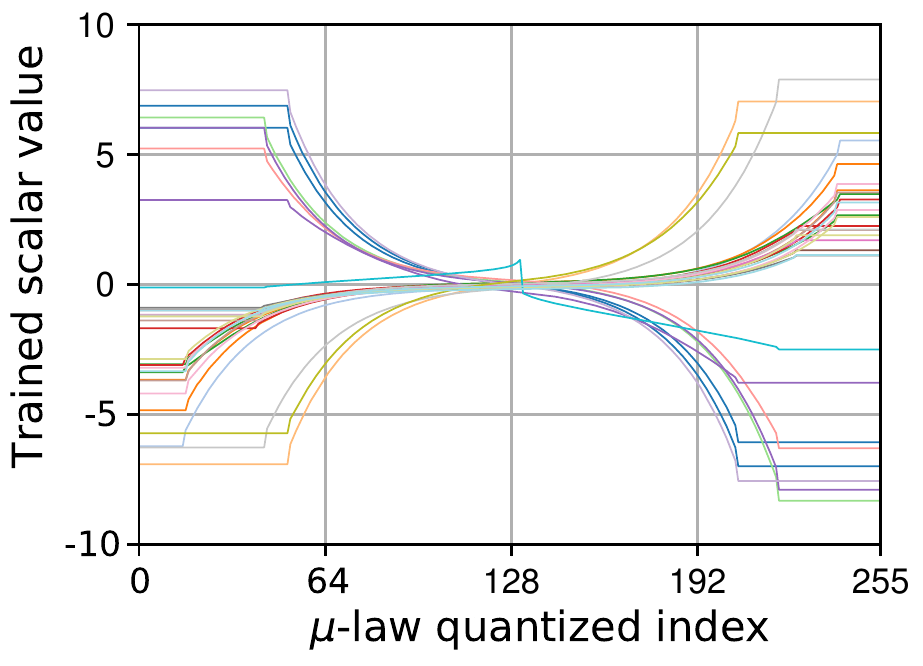}
\caption{The plot of trained embedding value with $n_{e}=1$.\\\hspace{\textwidth}(A total of 30 lines = 2 speakers $\cdot$ 3 feedback samples $\cdot$ 5 Sample Bunching)}
\label{fig:embed}
\end{figure}

Figure~\ref{fig:embed} illustrates the trained embedding values of \textit{B-LPCNet2-S}. Note that the horizontal lines at both ends arose from our optimization trick that maps the sparse indices to the adjacent dense index, preventing access to untrained weights in inference. Each line has a different scale, slope, symmetrical point, and nonlinearity. This suggests that the trained embedding table with $n_{e}=1$ controls the nonlinear transforms with desirable properties from the network. Evidently, this is more outstanding than directly feedbacking the scalar value.

\section{Conclusion}
\label{section:conclusion}
We introduced Bunched LPCNet2, which is a highly efficient neural vocoder architecture, and some presets to provide high-fidelity real-time TTS service to satisfy various device specifications, as confirmed by the MOS tests and RTF evaluations. We also investigated a simple technique that can significantly reduce the model size without degrading the speech quality. Compared with our previous study~\cite{bunched}, \textit{B-LPCNet2-R} achieved better speech quality and lower complexity with a model size of only 1.1 MB. In terms of quality, complexity, and model footprint, Bunched LPCNet2 would be the best candidate for low-resource edge devices, such as smart watches, wireless earphones, and home assistant devices.

\bibliographystyle{IEEEtran}

\bibliography{mybib}

\end{document}